# SOCIAL MEDIA MARKETING ANALYTICS: A MULTICULTURAL APPROACH APPLIED TO THE BEAUTY & COSMETICS SECTOR

Hajer Kefi
*National University of Singapore*, hajer.kefi@parisdescartes.fr

Sitesh Indra
*National University of Singapore*, sitesh.indra@telecom-paristech.fr

Talel Abdessalem
*National University of Singapore*, talel.abdessalem@telecom-paristech.fr

# SOCIAL MEDIA MARKETING ANALYTICS: A MULTICULTURAL APPROACH APPLIED TO THE BEAUTY & COSMETICS SECTOR

Hajer Kefi, CEDAG EA 1516, Université Paris Descartes; DRM CNRS, Université Paris Dauphine, France ; School of Computing, National University of Singapore, Singapore, hajer.kefi@parisdescartes.fr

Sitesh Indra, LTCI & IPAL, CNRS, Télécom ParisTech, Université Paris-Saclay, France ; School of Computing, National University of Singapore, Singapore, sitesh.indra@telecom-paristech.fr

Talel Abdessalem, LTCI & IPAL, CNRS, Télécom ParisTech, Université Paris-Saclay, France ; School of Computing, National University of Singapore, Singapore, talel.abdessalem@telecom-paristech.fr

Abstract

*We present in this paper a multicultural approach to social media marketing analytics, applied in two Facebook brand pages: French (individualistic culture, the country home of the brand) versus Saudi Arabian (collectivistic culture, one of its country hosts), which are published by an international beauty & cosmetics firm. Using social network analysis and content analysis, we identify the most popular posts and the most influential users within these two brand pages and highlight the different communities emerging from brand and users interactions. These communities seem to be culture-oriented when they are constructed around socialization branded posts and product-line oriented when advertising branded posts are concerned.*

# 1 INTRODUCTION

Recently, social media platforms have become an important channel for companies to deploy international marketing efforts in order to gain or maintain a global presence, especially by the means of brand (fan) pages. Previous research has provided limited insight into the best way for marketers to use social media (Burg, 2013; Stelzner, 2014), and even less is known to our knowledge about the cross-national effectiveness of social media marketing (Muk et al., 2014; Goodrich & Mooij, 2014). This research helps address this gap and uses a cross-cultural social network analysis to investigate a brand page belonging to an international French beauty & cosmetics company published globally on Facebook within more than 50 countries. In this paper, we present our approach and focus on the similarities and differences between this Facebook brand page deployed in France (an individualistic culture) and the one deployed in Saudi Arabia (a collectivistic one) (Hofstede, 2001; de Mooij and Hofstede, 2010). We argue here that it is important to examine and make sense of the data generated by marketers, users and the interactions between them, within an analytic canvas which can hardly be culture free when we are at an international setting. We use real data generated form all the activity accomplished by the brand proprietary and the users on the two fan pages during six months. We have chosen France because it is the home's brand and compare it to Saudi Arabia, which is not only very different in terms of cultural specificities but also a critical local market (in terms of turnover and competitive and strategic positioning in the Middle-East region). Moreover, two of the co-authors live in France, speak French and Arabic fluently and are familiar with the two cultures, which is very helpful for content analysis and cultural specificities decryption.

In the following sections, we first present a literature review on social media marketing and social network analysis, and their specificities in multicultural settings. Then, we present our framework for analysing social owned and social earned data (Stephen & Galak, 2012) related to the activities of respectively marketers and users on the brand pages under study. We finally discuss the implications, limitations and future steps of this research.

# 2 LITERATURE REVIEW

The interest devoted to value-based and cultural differences in marketing research is not new (Vinson et al., 1976; Segal et al., 1993). Globalization issues are often advocated to justify this interest. The divergences in social identity and value-expressive attitudes within our increasingly cosmopolitan societies are also at stake. Social media marketing, a new branch in marketing practices, seems to follow the same orientation toward a culture bounded approach to social media environments (De Mooij and Hofstede, 2010). These are considered as inherently linked to digitally-enabled social networks, i.e. "a complex assemblage of engagement, reflection, action, technology, organization and community" (Germonprez & Hovorka, 2013). They support and enable social media marketing actions, and are challenging for researchers from different fields, such as computer science, management, sociology, etc. In this section, we define social media marketing as a domain of research and practice. Then, we explain how social network analysis could help make sense of these complex environments and therefore how they could be applied in multicultural social media marketing.

## 2.1 Social Media Marketing and Cross-Cultural Research

The rapid growth in popularity of social media has exhorted marketers toward deploying their strategic campaigns within these platforms, especially on Facebook the most popular online social network in terms of number of users (1Facebook billion and 600 million, STATISTICA 2015) and also the most plebiscite by marketing practitioners (Stelzner, 2014). Social Media Marketing (SMM) can simply be defined as marketing in social media but the question about how specific it is within the marketing-mix arsenal and more precisely within digital marketing (marketing using the Internet channel) is still debated (Tiago & Verissimo, 2014). For Xie & Lee (2015), SMM has to be addressed within a multi-channel perspective of marketing actions because it is rarely deployed exclusively and/or separately from traditional marketing. Trusov et al. (2009) compare the effects of traditional marketing and online Word–Of-Mouth (e-WOM); Stephen & Galak (2012) consider jointly the effects of traditional and social media on sales; Rishika et al. (2013) investigated the consumers' participation activities in a firm's Facebook brand fan page, and argued that these activities are positively related to consumers' offline visit frequency. Furthermore, this multi-perspective view has helped identifying the specificities of SMM, which seem to be the following: (1) it is reputed to be less expensive than traditional marketing, (2) more likely to generate consumer's confidence, and (3) interactive (*versus* unidirectional for traditional marketing). A dominant tendency within this body of research seems to be dedicated to content production and sharing, and the interest has been first put on User-Generated-Content (Cui et al., 2012; Lipizzi et al., 2015), before pointing out the importance to jointly study consumer generated and marketer generated contents, and their interactions (Goh et al. 2013; Schulze et al. 2014; Chang et al., 2015; Homburg et al. 2015).

By asking: "how 'social' are social media" in their study of the cross-cultural differences between social media-based and offline purchase decisions, Goodrich & Mooij (2014) have demonstrated that culture cannot be ignored in social media marketing practices, for at least two reasons: (1) the way people use social media varies worldwide (Vasalou et al., 2010); and (2) cultural differences impact how people interact with products and brands whether online or via traditional shopping channels (Choi & Miracle, 2004; Goodrich & De Mooij, 2011). An evolving interest on cross-cultural SMM research has therefore emerged recently.

As reminded by De Mooij and Hofstede (2010), Hofstede's model (2001) has been very useful to address questions concerning global brand image, brand equity and advertising in cross-cultural contexts. Especially the bi-dimensional individualist/collectivist model, has been extensively used as a surrogate to address cultural differences in cross-cultural digital and social media marketing (De Mooij, 2010; Choi and Totten, 2012). For instance, Minton al. (2012) conducted a comparative analysis of motives for sustainable behaviors in individualistic (United States, Germany) *versus* Collectivistic (South Korea) cultures and have concluded that motives are complex, demanding careful analysis from advertisers who plan to deliver green advertisements over social media. For Goodrich and De Mooij (2014), relationship-oriented collectivists rely to a greater extent than individualists on social media, considered as an alternative for interpersonal WOM communication. And closer to our research issue: Muk et al. (2013) have applied a cross-national study of the influence of individualism *versus* collectivism on liking brand pages using an online study based on the theory of reasoned action (Ajzen & Fishbein, 1980).

Individualism/collectivism can be defined as "*people looking after themselves and their immediate family only, versus people belonging to in-groups that look after them in exchange for loyalty*" (De Mooij and Hofstede, 2010, p. 88-89). Following Okazaki and Taylor (2013), we argue that individualism/collectivism is perhaps the most central dimension of cultural variability identified in cross-cultural research and has to be considered in promoting a brand globally, e.g. do we need to emphasize persuasion for individualistic audiences *versus* creating trust for collectivistic ones, as advocated by de Mooij and Hofstede (2010)? In this paper, we investigate the hypotheses according to which content displayed in these pages and the degree of users' engagement toward this content may differ within the individualistic/collectivistic cultural dimension, namely in the French (deemed individualistic) *vs* Saudi Arabian (deemed collectivistic) brand fan pages.

**Social Network Analysis and Culture**

Social Network analysis (SNA) can be defined as a 'mode of inquiry' of social phenomena rooted in several theoretical perspectives, amidst them psychiatry (Moreno, 1934), anthropology (Barnes, 1954; Bott, 1957) and structural sociology (Blau, 1982; Wellman, 1981). The primary postulate of SNA is that "*the structure of social relations determine the content of those relations*" (Mizruchi, 1994, p. 330). Accordingly, SNA relies on the relationships between entities (individuals, organizations) rather than the attributes of those entities (socio-psycho & demographic for individuals, structural/institutional for organizations). In the last 20 years, SNA applications have rapidly grown in the economic and financial fields to study among others trade and financial networks (Boss, et al. 2004; Garlaschelli & Loffredo, 2005), to investigate inter-organizational networks (Nagurney et al. 2002; Nagurney & Qiang, 2009; Ying et al., 2011), and finally what is in concern in this article Internet and computer mediated social media and their implications in marketing and business intelligence.

According to Mizruchi (1994), the major theoretical achievements of SNA are twofold: the recognition of the effects of actor centrality in behaviour and the identification of network subgroups. Both have been enhanced by theoretical theses from sociology (Burt, 1992; Granovetter, 1985, etc.), confirmed by experimental and non-experimental studies and mathematically formalized into graphs.

From a SMM and business analytics perspective, SNA is extensively used for the following purposes: (1) *topological analysis* (to find the structural properties of a network or a graph, which is generally represented by a set of nodes connected by links); (2) *information flow analysis* (to determine the direction and strength of information flows through the network), (3) *centrality analysis* (to determine salient roles within a network), (4) *community analysis* (to find out clusters whose members are specifically connected), (5) block modeling (to discover key links between different clusters in a network), (6) *structural equivalence* (to identify network actors with similar specificities).

In the literature, the focus is mainly put on (1), (3) and (4), for example to build a blog mining and consumer interactions analysis approach (Chau & Xu, 2012), or to run a socio-semantic analysis of customers' reactions to the launch of new products (Lipizzi et al., 2015). More generally, because SMM is inherently network-based (Kaiser et al., 2013), SNA is a very helpful tool to be used. One of the challenging issues to be addressed could be to identify the most influential customers (with regard to their centrality in the network), and also to find out the cluster of network members which must be targeted in order to trigger an optimized wave of influence (Kalyanam et al., 2007).

An important stream of research is also devoted to viral marketing, social influence and contagion within diverse social media communities using experimentations and sophisticated data analytics designs including SNA (Aral & Walker, 2011; 2012). This has undoubtedly contributed to enrich the methodological toolkit that could be used in this research area.

Concerning the relationship between SNA and culture, from a sociologist point of view this relationship seems to be elusive. As explained by Lazega (1994; 2013), SNA is mainly focused on the relational specificities of social entities and recognizes (their) norms and values (the constituent elements of their culture) only implicitly and/or *a posteriori*. More emphatically, Emirbayer and Goodwin (1994) consider this relationship as theoretically under developed: "*gains its purchase upon social structure only at the considerable cost of losing its conceptual grasp upon culture, agency and process*" (Emirbayer and Goodwin, 1994, p. 1446-7). Now, from a practical and empirical point of view, SNA in multicultural settings has been for example applied in intercultural socialization studies (Kadushin, 2012; Chi & Suthers, 2015) because it has been recognized as useful to identify structural patterns that might be conducive or constraining for individuals' cultural adaptation such as ethnic diversity or cohesiveness of hosting communities.

Here again and following Okazaki and Taylor (2013), we can intuitively consider that social networks (user-user and user-brand interactions) built within brand pages could be related to individualism/collectivism because these related communities are not inclined to develop the same social patterns: loosely linked for individualist cultures *vs* in-group subordinate for collectivist ones, for example. We will therefore examine in this paper how these differences could be reflected in the SNA specificities (namely (1), (3) and (4)) in the two communities under study French *vs* Saudi Arabian.

# 3 THE CASE OF A BEAUTY & COSMETICS FACEBOOK BRAND PAGE DEPLOYED IN FRANCE AND SAUDI ARABIA

Social networking sites (SNS) initiated the possibility for companies to publish brand pages to interact and communicate with their brand communities (followers/fans). Facebook emerges as a leading actor in this area (Xie & Lee, 2015) with over 3 million active brand pages, with the top 20 exceeding each 20 million fans. Companies have the possibility to post branded content on their pages (including videos, brand messages, contests and other promotional materials like coupons) that can be used and shared by their fans (De Vries et al. 2012). Consumers can become brand page fans by clicking the "like" button on the brand page or on any of the brand Facebook ads. They can interact with the brand and/or with the other members of the community (De Vries et al., 2012). Statistics and ratings of Facebook brand pages in terms of audience (number of fans) are published by some SMM commercial platforms (e.g. http://www.socialbakers.com).

From the academic research perspective, only a few studies focus on assessing the effectiveness of brand pages, and use most of the time online surveys to investigate the determinant factors of users' (fans) engagement in various activities on these pages (Wolny & Muller, 2013; Muk et al., 2013; Smith, 2013). In another research avenue, practical recommendations are proposed concerning the configuration of the page, in order to drive more consumers' reactions (De Vries et al., 2012). We can nevertheless find only very few studies focusing on brand pages in the beauty & cosmetics industry. The study of Shen and Bissel (2013) is worth to be mentioned here, the authors initiated an exploratory investigation using content analysis of Facebook posts from six beauty brands to compare their branding strategies.

In this research, we present our framework for collecting, analysing and interpreting the data directly generated by all the activities within Facebook brand pages in order to produce social intelligence and actionable SMM knowledge. Our framework consists of the following steps:

1. *Brand page(s) Identification*: a firm can publish one page (for one brand) or many (for a range or a portfolio of brands, in one or many countries). The resulting fan pages may differ not only in terms of language but also concerning their configuration and content. To identify the pages we need to investigate is therefore determined by the level of marketing strategy we target (local, multi-local, global…). In this study, we have chosen two brand pages belonging to the same brand deployed in two countries: France, the brand home country; and Saudi Arabia, one of its hosts and market of a critical importance in the Middle-East and Arab region.

2. *Data Collection*: Because many brand pages have large audiences consisting of millions of fans (e.g. 94 015 245 for Coca Cola, http://www.socialbakers.com/, October 2015), we have to be able to handle big data crawling tools.

3. *Data Analysis*: concerns data generated by the brand page owner (or his/her SMM agents) and the users (fans). Here, we can differentiate between two types: social owned media and social earned media (Stephen & Galak, 2012).

    a. Social owned media includes all the data generated by the brand owner activity (or his/her agents) on the fan page, i.e. in the case of Facebook: the page owner's posts, comments, comments-replies, likes and shares.

    b. Social earned media corresponds to the data generated by the fans or users of the brand page, possibly in reaction to the page owner SMM actions. In the case of Facebook, this includes users' posts (when they are allowed to), comments, comments-replies, likes and shares.

    We investigate two sets of networks: the first represents the interactions between the page owner and the users and the second relies to the users-users interactions. Then, we conduct three major procedures of SNA: topological analysis, centrality analysis, and community analysis. Finally, we focus on data content, namely the textual content of all the posts, comments and comments-replies.

4. *Data Interpretation:* Based on our analyses, we aim to identify the most popular posts, especially those of the brand's owner (which could be part of a given SMM campaign), the most influential users (to be probably targeted in future actions), the different relevant communities and their specificities, etc. We also highlight the similarities and differences concerning all these aspects in our two fan pages.

In the remainder of this section, we first describe our data set and then our data analyses, results and interpretation.

## 3.1 Data Set

We have chosen one of the leading actors of the beauty & cosmetics industry in France (noted YX for confidentiality reasons). This brand has extensively engaged in digital marketing since the early 2000s and especially in SMM. YX is deploying Facebook brand pages in almost all the countries in which it sells its products (online and offline) all over the world. We are reporting in this study the case of YX Facebook brand page France which presents the largest audience of all the group (more than one million fans; the global number of all YX fan pages is slightly over two millions; ranked 5$^{th}$ in the beauty & cosmetics industry in France in terms of audience on Facebook (http://www.socialbakers.com/, 2015) and YX brand page Saudi Arabia.

An important specificity of YX Facebook brand pages has to be mentioned: Posts (containing videos, messages, quizzes, information, contests and other material) are only published by YX; who can also comment, reply on comments, share and like; whereas, users cannot publish posts. They can only react to YX and to other users' activities, by liking, sharing, commenting and replying to comments.

To extract the data, we used Netvizz: a Facebook API (Application Programming Interface) designed to extract data from Facebook pages. Netvizz is integrated directly into the platform. It allows the collection, extraction and exportation in standard file formats from different sections of the Facebook social networking service (Rieder, 2013). This API includes a search module, statistics for links shared on Facebook, networks of pages, group data and page data. Facebook is imposing a "fan-gate", so to access data and use Netvizz, we had to 'like' the pages we intend to explore. We have used the *page data* Netvizz module to extract data from YX fan pages. Data are here explored as "friendship" networks (or graphs). We have accessed the Fan pages on 09/12/2015 and extracted data related to all the pages' activity during the last six months (but not exceeding 999 posts: a limitation by Facebook). For each fan page, the application has provided us with three files: the first (data file 1) represents a bipartite network, where nodes are posts or users. When a user comments or likes a post, a directed edge between user and post is represented. The second (data file 2) is a tabular data file ready for statistical analyses. Finally, the third (data file 3) contains users' comments grouped per post, to facilitate content analysis. Table 1 shows the specificities of our data set (France & Saudi Arabia brand pages).

| Country | Language | Number of fans | Last post | Graph specificities | |
|---|---|---|---|---|---|
| YX France | French | 1 012 295 | 2h | Data extracted from 836 posts, with 30732 users liking or commenting 97182 times | 31568 nodes |
| | | | | | 90963 edges |
| YX Saudi Arabia | Arabic/English | 108244 | 7h | Extracted data from 175 posts, with 3357 users liking or commenting 8396 times | 3532 nodes |
| | | | | | 7425 edges |

*Table1: France and Saudi Arabia Facebook brand pages data set*

## 3.2 Network Analysis and Visualization

First of all, besides the bipartite graph provided by Netvizz (data file 1), we constructed a monopartite graph which allows us to process data concerning user-user interactions. It constitutes our data file 4. Then data (from data file 3 and data file 4) are imported into Gephi and processed. Gephi is an open and easy to use interactive platform for the visualization and exploration of networks, complex systems, dynamic and hierarchical graphs. For the French brand page, we particularly use Gephi Force Atlas2 a force-directed layout algorithm able to repulse nodes which are 'different' and at the same time attract similar ones, in order to express structural proximities into visual proximities (Network visualization), and also to facilitate social network analysis (Bastian et al., 2009).

We proceed to three of the major SNA analyses: topological analysis, centrality analysis and community analysis.

*Topological analysis* aims to determine the structural properties of the network. It can be assessed using the number of nodes and links, the density (ratio of the number of links in the network over the total number of links between all pairs of nodes), and the average path length (the average distance between two any nodes), etc.

*Centrality analysis* aims to determine the nodes which play important roles in the network. Centrality can be assessed using a range of statistics, such as degree, betweenness, closeness, Eigenvector, etc. (Freeman, 1979). In our case, nodes are either posts (by YX) or users. However, we have to mention that YX is considered in the graph as a user-node when he replies to users' comments (User-Id: page-owner in tables 3 and 4 below). To identify the popularity of these two categories, i.e. their connectedness to other nodes and hence their capacity to spread information, we will not use the same indicator. For instance, posts are popular when they are intensively commented, shared, liked, etc. (each of these activities represents an incoming link that leads into the post). Whereas, users are more or less influential, depending upon the intensity of their activities (outgoing links). So, we can calculate the weighted out-degree (the number of out-going links with regard to the weight of each link) for users; and the weighted in-degree for posts. Moreover, we calculate a closeness indicator for posts: Eigenvector centrality, which is a measure of reach (or speed) for a given post to attain any user.

Considering the specificities of our network (Opsah et al., 2010), we have chosen to rank users according to their weighted out-degree; and posts according to their Eigenvector centrality. Tables 2 and 3 present, respectively, the 10 most popular posts and the ten most influential users (posts-Ids, users-Ids and pages url are masked to preserve confidentiality) for the two pages under investigation (France *vs* Saudi Arabia).

*Community analysis* aims to identify within the network subgroups of nodes which ones are presenting denser links with each other than with nodes out of their subgroup (Wasserman & Faust, 1994). Each subgroup constitutes a cluster of nodes or a community. The procedure here is threefold: (1) *Community detection* to determine how many communities compose the network and the size of each of them (in terms of number of nodes); (2) *Modularity* which determines the strength of division of a network into communities. A high modularity implies dense connections between the nodes within a community but sparse connections between nodes belonging to different communities; (3) *Community interpretation* relates to the content exchanged. For example, inside a given community what are people talking about. In our case, the thematic content of the posts and the specificities of the users belonging to a given community are both useful for community interpretation.

In tables 2 and 3, we indicate the communities to which belong the most popular posts and users. In table 4, we present the 5 largest communities (representing 78,27% of the total nodes in the YX France graph; *vs* 92,76% in the YX Saudi Arabia). For each community, we go back to the posts which compose that community and especially to the most important of them. For this purpose, we use a page-rank procedure (a variant of Eigenvector centrality which displays the results in terms of url pages) to identify the top-ranked posts urls. We then examine (manually) the content information of each post and identify the thematic content of these communities.

*Community Interpretation*

Concerning YX brand page France, it appears that the communities ranked 1st and 5th (respectively (29) and (5)) are constructed around posts that have no advertising content. They could be described as 'socialization' posts aimed to create and sustain social links between the brand and the users, no products to advertise or sell are discussed here. Community (29) is constructed around a single post: Labour Day Greetings (the 1st of May, in France is very popular. A day in which French people, traditionally, offer friends and relatives a white flower called Lilly of the valley). Similarly, community (5) is constructed around a single post: "Happy Mother's day", which is locally very popular, too. The other communities are concerned with YX products. Community (2) is concerned with YX make-up products; community (40) is concerned with fragrances; and community (24) seems to be more interested by 'Botanical cosmetics' a leading YX product line comprising Hygiene and body care natural products. In these three communities, posts are mostly advertising-oriented and contain a link to the official website to shop online the products concerned. Figure 2 below shows how the communities are displayed by Gephi (to improve the visibility, we used here data concerning only 50 posts).

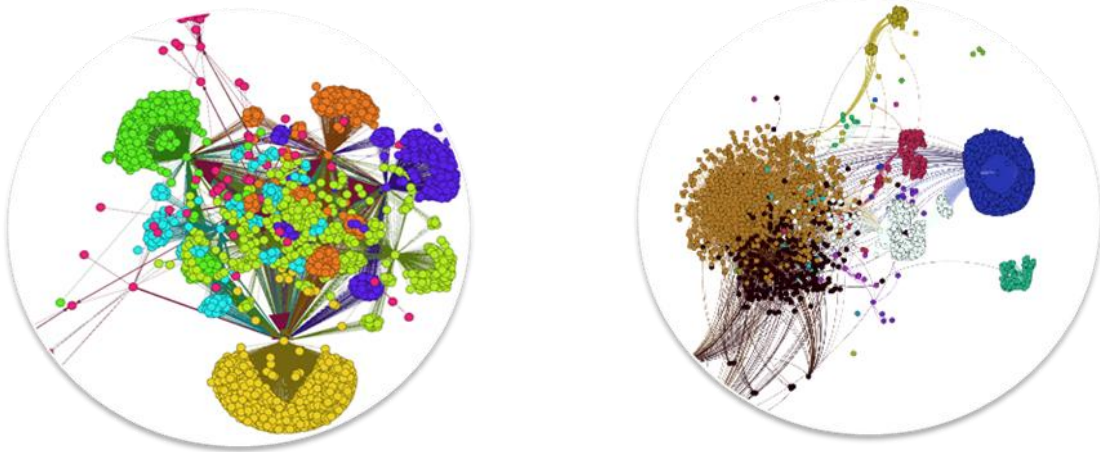

YX France brand page                                              YX Saudi Arabia brand page

(limited to the 50 last posts for a better visualization)

*Figure 1.        Community Detection: France vs Saudi Arabia*

| YX Fan page France | | | | YX Fan page Saudi Arabia | | | |
|---|---|---|---|---|---|---|---|
| Post_id | Type_post | Eigenvector Centrality | Community | Post_id | Type_post | Eigenvector Centrality | Community |
| 7151471151_####### | photo | 1 | 30 | 269780463061741_## | photo | 1 | 13 |
| 7151471151_####### | photo | 0,381985294 | 16 | 269780463061741_## | photo | 0,420072115 | 18 |
| 7151471151_####### | photo | 0,234283088 | 33 | 269780463061741_## | photo | 0,108173077 | 0 |
| 7151471151_####### | photo | 0,167647059 | 40 | 269780463061741_## | photo | 0,066105769 | 6 |
| 7151471151_###### | photo | 0,149816176 | 40 | 269780463061741_## | photo | 0,044471154 | 3 |
| 7151471151_###### | photo | 0,136764706 | 8 | 269780463061741_## | photo | 0,044471154 | 3 |

| 7151471151_###### | photo | 0,132904412 | 40 | 269780463061741_## | photo | 0,043269231 | 4 |
| 7151471151_###### | photo | 0,126470588 | 24 | 269780463061741_## | photo | 0,042067308 | 1 |
| 7151471151_###### | photo | 0,123805147 | 23 | 269780463061741_## | photo | 0,0390625 | 1 |
| 7151471151_###### | photo | 0,122150735 | 24 | 269780463061741_## | photo | 0,037259615 | 1 |

*Table 2: The 10 most popular posts ranked by their Eigenvector Centrality (France vs Saudi Arabia)*

| YX Fan page France | | | YX Fan page Saudi Arabia | | |
|---|---|---|---|---|---|
| User_Id | Weighted Out-Degree | Community | User_Id | Weighted Out-Degree | Community |
| Page owner | 1951 | 0 | user_##### | 150 | 1 |
| user_##### | 427 | 0 | user_##### | 136 | 1 |
| user_##### | 152 | 0 | user_##### | 122 | 1 |
| user_##### | 113 | 16 | user_##### | 108 | 1 |
| user_##### | 113 | 0 | user_##### | 106 | 1 |
| user_##### | 108 | 16 | user_##### | 104 | 1 |
| user_##### | 104 | 2 | user_##### | 103 | 0 |
| user_##### | 103 | 24 | Page owner | 96 | 4 |
| user_##### | 103 | 0 | user_##### | 94 | 1 |
| user_##### | 101 | 24 | user_##### | 81 | 1 |

*Table 3: The 10 most influential users ranked by their weighted out degree (France vs Saudi Arabia)*

| YX Fan page France | | | | YX Fan page Saudi Arabia | | | |
|---|---|---|---|---|---|---|---|
| Cluster_Id | Theme | Nb of posts | % Nodes | Cluster_Id | Theme | Nb of posts | % Nodes |
| 29 | 1st May Labour day | 1 | 23,71 | 13 | Beauty Arab eyes | 1 | 43,97% |
| 2 | Make-up (Link to shopping website) | 52 | 18,49 | 1 | Skin-care & fragrances | 114 | 17,64% |
| 40 | Fragrances (Link to shopping website) | 43 | 16,69 | 18 | Contests | 1 | 17,10% |
| 24 | Botanical Cosmetics (Link to the shopping website) | 29 | 12,29% | 4 | Make-up products | 38 | 8,61% |
| 5 | Mother's day | 1 | 7,09 | 0 | Beauty Arab eyes (smoky eyes) | 2 | 5,35% |
| Total 5 largest clusters (% Nodes) | | | 78,27 | Total 5 largest clusters (% Nodes) | | | 92,76 |

*Table 4: Clusters Interpretation (France vs Saudi Arabia)*

For YX brand page Saudi Arabia, the largest community (13), representing 43, 97% of the total nodes, is constructed around a single post 'Arab Beauty Eyes' which is a specific content developed only in the Saudi Arabian fan page (and some other Arab countries). This can be considered as a socialization post with no advertising content. The 5[th] community (0) is also concerned by beauty eyes but it is constructed around an advertising content ('Smoky eyes', which is a global advertising' campaign deployed in all YX brand pages). Communities (1) and (4) concern a specific interest in some of YX product lines; respectively skin-care/fragrances and make-up products. Brand-users' and users-users' interactions are here about the physical stores locations and prices, no links to shopping websites are provided. Finally, community (18) is constructed around a contest. Posts and comments are in English or in Arabic, while products' names are in French and some of the advertising contents refer to Paris in order to emphasize YX brand home.

We have to mention here is that 'big smoky dark eyes' represents one of the most salient female aesthetic ideals in the Arab culture, jointly with an amber fragrance and a lightened skin. Moreover, in the Muslim tradition (where women have to be veiled), it is allowed to highlight the face and especially the eyes. These cultural specificities seem to be very helpful to interpret the community configuration of the Saudi Arabian YX brand page, compared to the French one. It seems also that YX is more engaged in the French brand page (ranked 1$^{st}$), compared to the Saudi Arabian one (ranked 8$^{th}$). More importantly, none of YX video posts has emerged within the top ten posts (neither in the top 20) in the two brand pages. If we consider the importance of videos to advertise beauty & cosmetics brands in the web, especially in the blogosphere, this can be considered as a critical shortage in YX's SMM efforts.

## 4 DISCUSSION AND CONCLUSION

We have constructed a framework to extract and analyse social owned and social earned data from two Facebook brand pages belonging to a French Beauty & Cosmetics firm. With regard to our findings, firms cannot be encouraged to standardize social media usage across borders, as it has been outlined by previous research (Berthon et al. 2012; Okazaki & Taylor 2013). In our case, it appears that YX is publishing global and culture specific content on its French and Saudi Arabian Facebook brand pages and that these communities react to these two types of content differently. Our content analysis shows also that the branded content is argumentation-oriented (defining the products specificities) in the French brand page and more visual and ludic (contests) in the Saudi Arabian. This globally corresponds to previous research concerning advertising design specificities recommended in individualistic *versus* collectivistic cultures (Okazaki & Taylor 2013; Bianchi & Andrews 2015). For each brand page investigated, our framework has provided clear insights to identify the most successful product lines, the most popular posts and who to target in order to foster the advertising campaigns (the most influential users). This framework can be applied in all local brand pages published by a firm and could also be extended to other firms for the purpose of an industry-specific competitive analysis. This represents an important managerial implication of our study. For instance, we are able to help marketers assess the effectiveness of their SMM actions and also improve the content they are deploying on their diverse Facebook brand pages.

Some of the choices and assumptions we have made in this research could be theoretically questionable, concerning for example the centrality indicators we are using, hence are they part of the limitations of this study, along with the use of manual content analysis. To overtake these limitations and go a step further with our research, we are planning to use automatic content analysis, sentiment analysis and text-mining (He et al., 2013) in larger multicultural brand pages samples within other industries.

Our approach is innovative, in that it uses social network analysis and content analysis to glean business intelligence from Facebook brand fan pages in multicultural settings. It brings therefore a valuable contribution to the research on social media marketing and analytics. We advocate in our study for the use of big data analytics in management information systems (Agarwal & Dhar, 2014) to exploit real data generated from web activities, and also for the importance to take into account cultural differences which are inherently embedded within these data, as they reflect human activity. This importance becomes vital when it concerns businesses related to culture-bounded products and services, as it the case of the beauty and cosmetics industry (Preiss, 1998), mainly because they involve self-representation and are highly sensitive in terms of aesthetic, social and moral norms.